\documentclass[fleqn,twoside]{article}
\usepackage{espcrc2}
\newif\ifpdf
\ifx\pdfoutput\undefined\pdffalse\else\pdfoutput=1\pdftrue\fi
\ifpdf
  \usepackage[backref,citecolor=blue]{hyperref}
  \let\myhref=\href\def\href#1#2{\penalty-20\myhref{#1}{\tt #2}}%
\else%
  \def\href#1#2{{\penalty-20\tt #2}}
\fi
\def\epspdffile#1{\leavevmode\ifpdf\epsffile{#1.pdf}\else\epsffile{#1.eps}\fi}
\usepackage{epsfig}
\font\tenbb=msbm10
\font\sevenbb=msbm7
  \newfam\bbfam
  \def\bb{\fam\bbfam\tenbb}
  \textfont\bbfam=\tenbb
  \scriptfont\bbfam=\sevenbb
  \scriptscriptfont\bbfam=\scriptfont\bbfam
\def\min{\mathop{\rm min}}		
\def\max{\mathop{\rm max}}		
\def\det{\mathop{\rm det}}		
\def\gcd{\mathop{\rm gcd}}		
\def\ln{\mathop{\rm ln}}		
\def\sgn{\mathop{\rm sgn}}		
\def\sn{\mathop{\rm sn}}		
\def\cn{\mathop{\rm cn}}		
\def\dn{\mathop{\rm dn}}		
\def\K{\mathop{\rm K}}			
\def\defn{\equiv}			
\def\Dslash{\hbox{D}\kern-0.6em\raise0.15ex\hbox{/}} 
\def\rational#1#2{{\mathchoice{\textstyle{#1\over#2}}%
  {\scriptstyle{#1\over#2}}{\scriptscriptstyle{#1\over#2}}{#1/#2}}}
\def\half{\rational12}                      
\def\N{{\bb N}}                     

\begin{document}

\title{Approximation Theory for Matrices}

\author{A. D. Kennedy\address{School of Physics, University of Edinburgh,
 King's Buildings, Edinburgh, EH9 3JZ, United Kingdom}}

\begin{abstract}
  {\noindent We review the theory  of optimal polynomial and rational Chebyshev
  approximations, and Zolotarev's formula for  the sign function over the range
  \(\epsilon\leq|z|  \leq1\). We  explain  how rational  approximations can  be
  applied  to  large sparse  matrices  efficiently  by  making use  of  partial
  fraction       expansions       and       multi-shift      Krylov       space
  solvers.\parfillskip=0pt\par}
\end{abstract}

\maketitle

\section{Introduction}

There are many situations in which it  is desirable to evaluate a function of a
matrix. For instance, in lattice quantum field theory it is sometimes desirable
to  evaluate the square  root of  a discretised  Dirac operator  \(\Dslash\) in
order  to calculate the  effects of  varying the  number of  fermionic flavours
\cite{forcrand97a,jansen97a,kennedy98a,Takaishi:2000je,Clark:2003na},   or   to
construct  a good  approximation  to Neuberger's  operator for  Ginsparg-Wilson
fermions         by         evaluating         the        sign         function
\cite{Neuberger:1997fp,Neuberger:1998my,Edwards:1998yw,Borici:2002a}.

\section{Matrix Functions}

As  a general  mathematical  problem we  need to  define  what we  mean by  the
generalisation of  a scalar function to a  corresponding matrix-valued function
on matrices. This  can be done in a variety of  ways \cite{golub:1996}, but for
our purposes  it suffices to restrict our  attention to the special  case of an
Hermitian matrix \(H\),  which can be transformed into real  diagonal form by a
unitary transformation\footnote{Although the lattice Dirac operator \(\Dslash\)
is  not  hermitian the  assumption  of  \(\gamma_5\)  hermiticity implies  that
\(\Dslash\gamma_5\)  is,  and this  may  be  used instead.}  \(H=UDU^\dagger\).
Defining a function  of a real diagonal  matrix to be a diagonal  matrix in the
obvious  way,  \(f(D)_{ii}  \defn  f(D_{ii})\),  we  may  define  \(f(H)  \defn
Uf(D)U^\dagger\).

We next address the question of how to compute good numerical approximations to
\(f(H)\) cheaply for a  useful class of functions. One way of  doing this is to
find a  polynomial or rational  function which is  a good approximation  over a
compact interval \(R\) containing the spectrum of \(H\). This seems a promising
approach because it is well known that rational approximations are an effective
way    of    computing    scalar    functions    on    a    digital    computer
\cite{rivlin:1969,achieser:1956,petrushev:1987,cheney:1966}.

The reason  why matrix polynomials  are cheap to  evaluate is that they  do not
require explicit diagonalisation of the matrix. From the identities
\begin{eqnarray*}
  U(\alpha D+\beta D')U^\dagger &=& \alpha UDU^\dagger + \beta UD'U^\dagger, \\
  UDD'U^\dagger &=& UDU^\dagger UD'U^\dagger,
\end{eqnarray*}
we  see that  \(p(H) =  p(UDU^\dagger)  = Up(D)U^\dagger\)  for any  polynomial
\(p\). Similarly, if we  observe that \(UD^{-1}U^\dagger = [UDU^\dagger]^{-1}\)
then we  see that \(r(H) =  r(UDU^\dagger) = Ur(D)U^\dagger\)  for any rational
function \(r\),  which is cheap to evaluate  if we count matrix  inversion as a
``cheap'' operation.

It is reasonable  to ask why we should consider  matrix inversion as ``cheap.''
There are three practical reasons for this
\begin{enumerate}
\item In applications we do not need a function of a matrix or the inverse of a
matrix  \emph{per se},  but only  the  effect of  applying it  to some  vector.
Therefore we  only need  to solve a  system of  linear equations, which  is far
cheaper than finding the full inverse.
\item  There are very  efficient methods  for solving  large systems  of sparse
linear equations. We may  take advantage of all the work that  has been done in
developing and optimising such Krylov space algorithms.
\item   We   can   expand    a   rational   function   in   partial   fractions
\cite{Neuberger:1998my,Edwards:1998yw,Chiu:2003ub}, and  then take advantage of
the structure of  Krylov space solvers to compute all the  terms in the partial
fraction       expansion      using       the      same       Krylov      space
\cite{Frommer:1995ik,Jegerlehner:1996pm}. For this to work it is necessary that
the partial  fraction expansion is numerically  stable, i.e., the  terms do not
suffer large cancellations;  and that the denominator of  the rational function
is  square-free.  For reasons  that  are not  fully  understood  both of  these
conditions seem to be satisfied in practice.
\end{enumerate}

Observe that the  definition of \(f(H)\) only requires that  we know the values
that  \(f\)  takes on  the  eigenvalues of  \(H\),  whereas  the polynomial  or
rational approximation must be valid  over an interval containing the spectrum.
For  certain functions,  such  as \(1/x\),  we  can find  cheaper methods:  for
instance, Krylov space  methods are usually much better  for finding an inverse
than using a polynomial approximation to \(1/x\) \cite{luescher94a}.

\section{Approximation by Polynomials}

\subsection{Function Norms}

In order  to define  more precisely  what we mean  by a  ``good'' approximation
\(p\) to  a continuous function  \(f\) over the  unit interval \([0,1]\)  it is
useful to introduce a family of norms, \[\|f\|_n \defn \left(\int_0^1 dx\, w(x)
\bigl|f(x)\bigr|^n\right)^{1/n},\]  where \(w\) is  a positive  weight function
and \(n\geq1\). We may verify that this  does indeed define a norm on the space
of continuous functions over the unit interval by the use of the Cauchy-Schwarz
inequality.  Some cases  of  especial interest  are  when \(w(x)=1\)  (absolute
norm),  \(w(x) =  1/|f(x)|\)  (relative norm),  \(n=2\)  (Euclidean norm),  and
\(n=\infty\) (Chebyshev  norm) for  which \[\|f\|_\infty =  \max_{0\leq x\leq1}
w(x) \bigl|f(x)\bigr|.\] In this last case if we choose the function \(p\) from
some  special  class of  continuous  functions,  like  polynomials or  rational
functions,  then the \(p\)  which minimises  the Chebyshev  norm of  \(p-f\) is
called the \emph{minimax}  approximation, \[\|p-f\|_\infty = \min_p \max_{0\leq
x\leq1} w(x) \bigl|p(x)- f(x)\bigr|.\]

If we are using floating-point numbers to represent components of spinor fields
on a computer then it is most  consistent to use a minimax approximation with a
relative  weight  for  matrix  function  evaluation. In  particular,  using  an
approximation with a minimax error  of one \emph{ulp} (unit of least precision)
means that  the errors caused  by function approximation  have no reason  to be
worse than those we already accept by using floating-point arithmetic.

\subsection{Weierstrass' theorem}

The  fundamental  theorem  on  the  approximation of  continuous  functions  by
polynomials is due to Weierstrass,  who proved that any continuous function can
be arbitrarily well  approximated over the unit interval  by a polynomial. This
theorem  is  of significance  in  functional analysis,  as  it  shows that  the
polynomials are dense in the space  of continuous functions with respect to the
topology  induced by  the  Chebyshev  norm. An  elegant  proof of  Weierstrass'
theorem may  be obtained  by the use  of Bernstein Polynomials,  \[p_n(x) \defn
\sum_{k=0}^n f\!\left(\frac kn\right) \left(\begin{array}{c} n \\ k \end{array}
\right) x^n (1-x)^{n-k}.\] It is  simple to show that \(\lim_{n\to\infty} \|p_n
- f\|_\infty = 0\).

\section{Optimal Approximation}

Weierstrass' theorem  tells us that for any  specified tolerance \(\epsilon>0\)
we can find  a degree \(n\) such that there is  a polynomial \(p_n\) satisfying
\(\|p_n - f\|_\infty < \epsilon\). For our purposes the converse question is of
greater importance, namely for a given  degree \(n\) what is the smallest error
\(\|p_n -  f\|_\infty\) that  can be achieved.  This question was  addressed by
Chebyshev, who provided a suprisingly simple answer. For pedagogical reasons we
shall  first discuss  Chebyshev's  theorem  for the  case  of approximation  by
polynomials, and  then generalise the  result to rational functions  (ratios of
polynomials); although logically the polynomial  case is just a special case of
the rational one  where the degree of the denominator  polynomial happens to be
zero.

\subsection{Polynomials}

Chebyshev  proved  that  for  any   degree  \(d\)  there  is  always  a  unique
polynomial\footnote{The  polynomial  may be  of  degree  less  than \(d\);  for
example  the best  approximation  of any  degree  to a  constant  is just  that
constant.  A less  trivial example  is  that the  zero polynomial  is the  best
approximation to  \(\sin N\pi x\) over  \([0,1]\), \(N\in\N\), for  all \(d\leq
N\).} \(p\) that minimises \(\|e\|_\infty = \max_{0\leq x\leq1} |e(x)|\), where
the error is \(e(x) \defn w(x)  [p(x) - f(x)]\). Furthermore this polynomial is
characterised  by the  criterion  that  the error  \(e(x)\)  takes its  maximum
absolute  value at  at least  \(d+2\) points  on the  unit interval  (which may
include the end  points of the interval), and the sign  of the error alternates
between its successive extrema.
\begin{figure}
  \epsfxsize=0.5\textwidth
  \centerline{\epspdffile{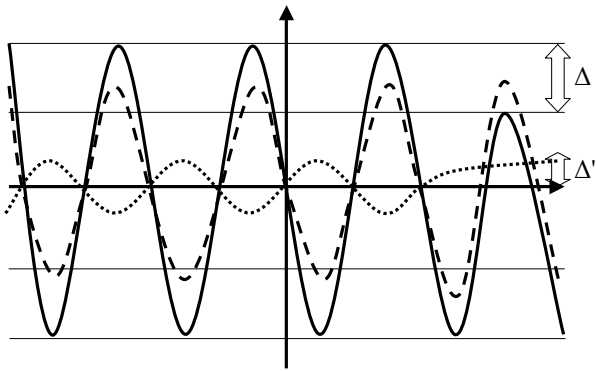}}
  \caption[Chebyshev  necessity]{The solid  line shows  the error  \(e\)  for a
  polynomial  approximation \(p\) of  degree \(d\)  to the  continuous function
  \(f\), which has less than \(d+2\) extrema of equal magnitude. This defines a
  ``gap'' \(\Delta\) between the extrema. We may construct the polynomial \(q\)
  of degree \(d\) shown by the dotted  line that has opposite sign to the error
  at  each  of  the   extrema,  and  whose  magnitude  \(\Delta'<\Delta\).  The
  polynomial \(p+q\) is  then a better approximation than  \(p\), as is obvious
  from its error \(w(x)[p(x)+q(x)-f(x)]\) shown by the dashed line.}
  \label{fig:chebyshev-necessity}
\end{figure}

\begin{figure}
  \epsfxsize=0.5\textwidth
  \centerline{\epspdffile{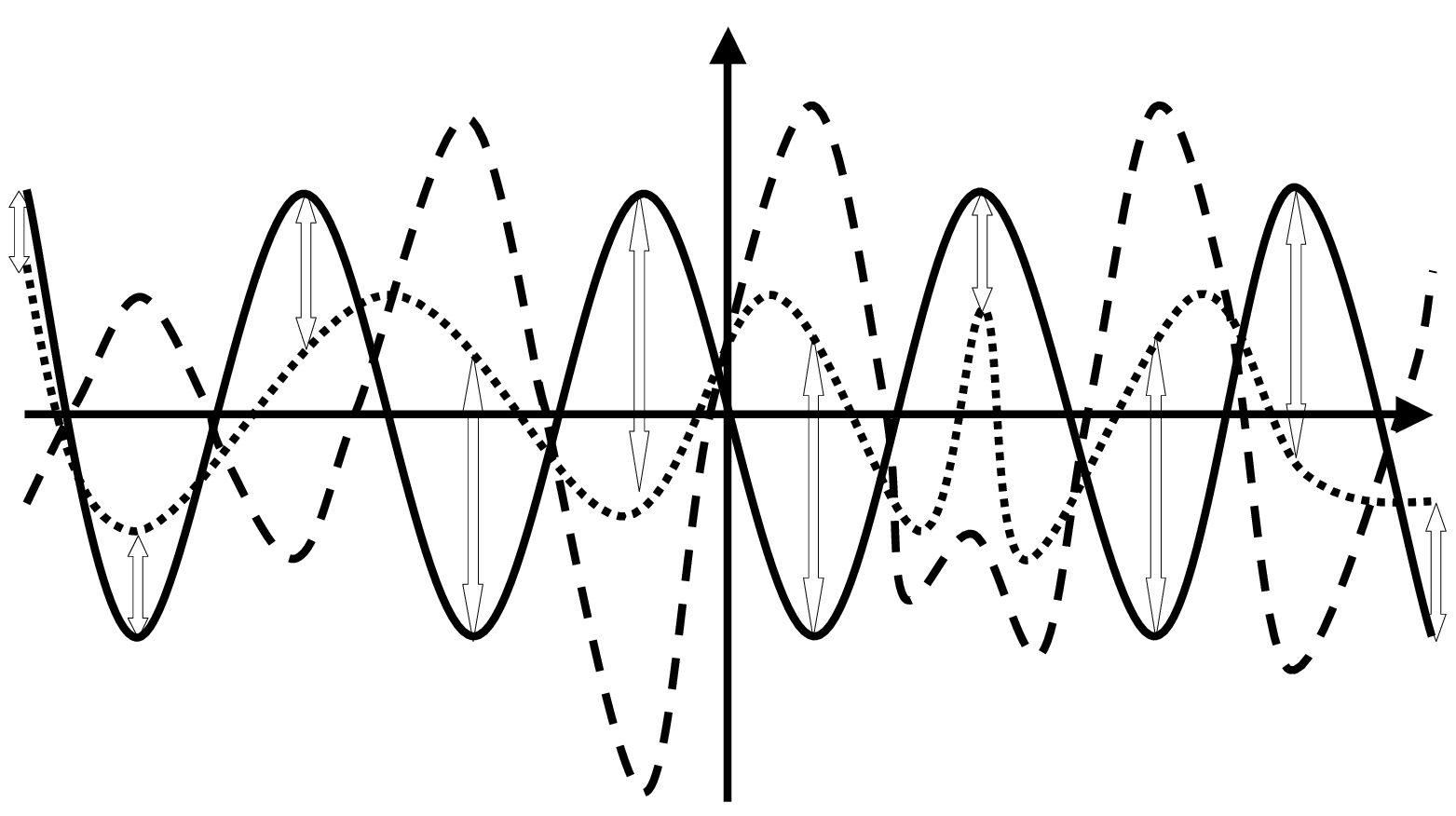}}
  \caption[Chebyshev sufficiency]{The  solid curve is  the error \(e_p\)  for a
    polynomial  \(p\) of  degree  \(d\) satisfying  Chebyshev's criterion.  The
    dotted  line   shows  the  error  \(e_{p'}\)  for   a  hypothetical  better
    approximation \(p'\). The dashed line is the polynomial \(p'-p\) which must
    have at least \(d+1\) zeros and therefore vanishes identically.}
  \label{fig:chebyshev-sufficiency}
\end{figure}

The  proof of  Chebyshev's theorem  is straightforward,  and is  illustrated in
Figures~\ref{fig:chebyshev-necessity}      and~\ref{fig:chebyshev-sufficiency}.
First we  prove that Chebyshev's criterion  is necessary: if  the error attains
its  extreme  value at  fewer  than  \(d+2\)  ``alternating'' points  then  the
approximation can be improved. Consider  a polynomial \(p\) for which the error
\(e(x) \defn w(x)[p(x)-f(x)]\)  takes its extreme values at  fewer than \(d+2\)
points (shown by the  solid curve in Figure~\ref{fig:chebyshev-necessity}); the
next largest extremum of the error (which  may be a local extremum or may occur
on the  boundary) must be smaller  by some non-zero ``gap''  \(\Delta\). As the
\(d+1\) extrema alternate in sign the error must pass through zero between each
successive extremum, and  there are at most \(d\) such  zeros \(z_i\) (if there
are several zeros  between adjacent extrema we just chose one  of them). We can
construct a polynomial  \(u\) of degree \(d\) (shown by the  dotted line in the
figure) which is zero at each of the \(z_i\). This may be written explicitly as
\(u(x) \defn A\prod_i (x-z_i)\). The constant \(A\) may be chosen such that the
sign of  \(u\) is  opposite to  that of the  error at  each extremuum,  and its
magnitude \(\Delta' = \|u\|_\infty =  \max_{0\leq x\leq1} w(x) |u(x)|\) is less
than the  gap \(\Delta\). It  follows that the  polynomial \(p+u\) is  a better
approximation to \(f\), as its error \(p(x) + u(x) - f(x)\) (the dashed line in
the figure) is everywhere less than \(e(x)\).

Next we  prove the sufficiency of  Chebyshev's criterion: if  the error attains
its extreme values at exactly \(d+2\)  alternating points then it is indeed the
optimal approximation. To  show this assume that there  was a polynomial \(p'\)
that    furnished    a    better    approximation.   The    solid    line    in
Figure~\ref{fig:chebyshev-sufficiency}   shows   the   error   \(e_p(x)   \defn
w(x)[p(x)-f(x)]\) for a polynomial  \(p\) satisfying Chebyshev's criterion. The
dotted  line shows  the error  \(e_{p'}(x)  \defn w(x)  [p'(x)-f(x)]\) for  the
hypothetical   better  approximation   \(p'\).   Since  \(p'\)   is  a   better
approximation   \(\|e_{p'}\|_\infty   <   \|e_p\|_\infty\),   or   equivalently
\(|e_{p'}(x_i)|  < |e_p(x_i)|\)  at  each  of the  \(d+2\)  extrema \(x_i\)  of
\(e_p(x)\). Appealing  to continuity we deduce that  \(e_{p'}(z_i) = e_p(z_i)\)
at at  least \(d+1\) points \(z_i\)  between the extrema. From  this it follows
that \(e_{p'}(z_i)  - e_p(z_i) = w(z_i)  [p'(z_i) - f(z_i)] -  w(z_i) [p(z_i) -
f(z_i)] =  w(z_i)[p'(z_i) - p(z_i)] =  0\), so the polynomial  \(p'(x) - p(x)\)
(shown by  the dashed line) must  have a least \(d+1\)  zeros, but as  it is of
degree \(d\)  this means it is  identically zero by the  fundamental theorem of
arithmetic.  Hence \(p'  = p\)  and the  polynomial \(p\)  is both  optimal and
unique.

\subsection{Chebyshev Polynomials}

The  Chebyshev   polynomials\footnote{It  is  easy  to  show   that  these  are
polynomials,     indeed    \[T_n(x)    =     \sum_{k=0}^{\lfloor    n/2\rfloor}
\left(\begin{array}{c}    n    \\    2k   \end{array}\right)    \sum_{\ell=0}^k
\left(\begin{array}{c}  k \\  \ell \end{array}\right)  (-1)^\ell x^{n-2\ell}.\]
The leading coefficient  of \(T_n(x)\), that is the  coefficient of \(x^n\), is
thus seen  to be \(\sum_{k=0}^{\lfloor n/2\rfloor}  \left(\begin{array}{c} n \\
2k  \end{array}\right)  = 2^{n-1}\).}  \(T_n(\cos\theta)  \defn \cos  n\theta\)
provide  a  simple  example   of  optimal  minimax  approximations.  \(T_n(x)\)
obviously has \(n+1\) extrema which alternate in value between \(-1\) and \(1\)
for \(-1\leq x\leq1\);  therefore \(p_n(x) \defn x^n -  2^{1-n} T_n(x)\) is the
best  polynomial approximation  of degree  \(n-1\) with  uniform weight  to the
function \(x^n\)  over the  interval \([-1,1]\), for  the error  \(e_n(x) \defn
p_n(x) - x^n = 2^{1-n}  T_n(x)\) satisfies Chebyshev's criterion. The magnitude
of the error is just \(\|e_n\|_\infty = 2^{1-n} = 2e^{-n\ln2}\), so we see that
in this case the error decreases exponentially with respect to~\(n\).

If we consider  a function \(f\) that has a  convergent Taylor series expansion
over \([-1,1]\), \[f(x) = \sum_{j=0}^\infty \frac{f^{(j)}(0)}{j!} x^j,\] and we
reexpress  it  as  an   expansion  in  Chebyshev  polynomials\footnote{This  is
straightforward  because they  satisfy the  orthogonality  relation \[\frac2\pi
\int_{-1}^1 \frac{dx}{\sqrt{1-x^2}} T_n(x)  T_m(x) = \delta_{n,m}\] (apart from
the trivial  case where \(n=m=0\)  for which there  is an additional  factor of
two).} \[f(x)  = \sum_{j=0}^\infty c_j T_j(x)\] then  truncating this expansion
at order \(d\) gives an approximation  close to the optimal one of this degree.
It  is in fact  the optimal  polynomial approximation  to the  truncated Taylor
expansion  of  degree \(d+1\),  which  is  not  necessarily \emph{exactly}  the
optimal  one. Likewise, the  convergence of  the truncated  Chebyshev expansion
\emph{may} converge  exponentially in  \(d\), but it  will not if  the original
Taylor series expansion did not converge exponentially.

\subsection{Rational Functions}

It is  perhaps surprising that Chebyshev's  argument can be  easily extended to
the  case  of  optimal  rational  approximations  as  well.  The  statement  of
Chebyshev's theorem  in this  case is  that for any  degree \((n,d)\)  there is
always  a  unique  rational  function  \(r\) that  minimises  \(\|e\|_\infty  =
\max_{0\leq  x\leq1} |e(x)|\), where  the error  is \(e(x)  \defn w(x)  [r(x) -
f(x)]\). For  rational approximations Chebyshev's  crtierion is that  the error
\(e(x)\) takes its  maximum absolute value at at least  \(n+d+2\) points on the
unit interval (which may include the  end points of the interval), and the sign
of the error alternates between its successive extrema.

The  proof  is similar  to  that  for polynomials  given  above,  and we  shall
therefore  just  briefly  consider  the  salient  differences.  To  prove  that
Chebyshev's  criterion is  necessary we  consider a  degree  \((n,d)\) rational
function \(r=p/q\)  for which the error \(e  \defn r - f\)  attains its extreme
value  \(\|e\|_\infty\)  at  fewer   than  \(n+d+2\)  alternating  points.  The
denominator  \(q\) cannot  have  any zeros  on  the unit  interval, as  clearly
\(q=1\)  would  then  lead  to  a  smaller error;  therefore  without  loss  of
generality we can assume \(q(x) > \delta > 0\) throughout the interval. Just as
before we may construct the polynomial  \(u\) of degree \(n+d\) that shares the
zeros of the \(e\). Since \(\gcd(p,q)\neq0\) we can construct polynomials \(s\)
and  \(t\) of degrees  \(d\) and  \(n\) such  that \(u  = sp  + tq\)  using the
extended Euclidean algorithm.  We then define the rational  function \(r' \defn
\frac{p-\epsilon t}{q+\epsilon s}\) which satisfies
\begin{eqnarray*}
  \lefteqn{|r'-f| = \left|\frac{p-\epsilon t}{q+\epsilon s} 
    - \frac pq + \frac pq - f\right|} && \\
  && = \left|- \frac{\epsilon(qt +  ps)}{q(q+\epsilon  s)} +  e\right|
     = \left|e - \frac{\epsilon u}{q(q+\epsilon s)}\right|.
\end{eqnarray*}
Choosing the constant  \(\epsilon\) small enough that \(q(x)  + \epsilon s(x) >
\delta'    >    0\)   for    \(x\in[0,1]\)    and   \(\epsilon\|u\|_\infty    <
\Delta\delta\delta'\) the required result follows.

The proof of sufficiency in the  rational case is even simpler. If \(r=p/q\) is
a degree \((n,d)\) rational  approximation whose error \(w(r-f)\) has \(n+d+2\)
alternating  extrema of  equal magnitude,  and \(r'=p'/q'\)  is  a hypothetical
better rational  approximation of  the same degree,  then the error  for \(r'\)
must cross that for \(r\) at least \(n+d+1\) times, i.e., the rational function
\(r-r'\) must have more than \(n+d\) zeros. Now, in order for \[r-r' = \frac pq
- \frac{p'}{q'}  = \frac{pq'-p'q}{qq'}\]  to vanish  on the  unit  interval the
numerator must  vanish (the denominator  certainly cannot); and  the polynomial
\(pq'-p'q\) is of  degree \(n+d\) and thus can only have  \(n+d+1\) zeros if it
vanishes identically, hence \(r=r'\).

\section{Remez Algorithm} \label{sec:remez}

Chebyshev's theorem  assures us that  there is a unique\footnote{In  some cases
the optimal  approximation may actually  be of lower degree.}  optimal rational
approximation of any degree to  any continuous function over the unit interval,
and it even provides a simple criterion  for the behaviour of the error in this
case, which is certainly of practical  use in verifying that a putative minimax
solution is correct.  In fact, the proof of the  theorem is quite constructive,
but it does not provide a particularly efficient means of computing the optimal
solution in practice.

In general, the computation of  the optimal Chebyshev rational approximation is
often performed using the \emph{Remez algorithm} \cite{remez:1957,cheney:1966}.
To describe the Remez algorithm we  define a \emph{reference} \(X\) as a set of
points \(\{x_i\in[0,1], i=1,\ldots,n+d+2\}\) that  we will make converge to the
alternating  set  at  which the  error  takes  its  extreme values.  The  Remez
algorithm alternates two steps:
\begin{enumerate}
\item  \emph{New  \(r\).}  Keep  \(X\)  fixed, and  choose  the  best  rational
approximation  \(r\) which  passes  through the  points  \(\left(x_i, f(x_i)  +
(-1)^i \Delta\right)\) for \(x_i\in X\).
\item \emph{New \(X\).}  Keep \(r\) fixed, and choose a  new reference which is
the best alternating set for \(r\) near to \(X\).
\end{enumerate}

In slightly more detail:
\begin{enumerate}
\item Interpolate an \((n,d)\) rational  function \(r=p/q\) such that the error
\(e(x_i) \defn w(x_i) [f(x_i) -  r(x_i)] = (-1)^i\Delta\). Setting \(p(x) \defn
\sum_{j=0}^n p_j x^j\)  and \(q(x) \defn \sum_{j=0}^d q_j  x^j\) with \(q_d=1\)
this  gives  the  \(n+d+2\)   equations  \(\sum_{j=0}^d  q_j  x_i^j  [f(x_i)  -
(-1)^i\Delta/w(x_i)] - \sum_{j=0}^n p_j x_i^j = 0\) for the \(n+d+2\) variables
\(p_j\), \(q_j\), and \(\Delta\). Noting that these equations are linear in the
\(p_j\) and \(q_j\), we write this  as the matrix equation \(Mv=0\) where \(v\)
is  the  vector  \((p_0,\ldots,p_n,q_0,\ldots,q_n)\).  This has  a  non-trivial
solution for \(v\)  iff \(\det M=0\), which is a  polynomial in \(\Delta\). For
the values  of \(\Delta\) which are real  roots of this polynomial  we can find
the coefficients  \(p_j\) and  \(q_j\), and check  that the  resulting rational
function \(r\) is valid.
\item Choose a partition of \([0,1]\) into intervals \(I_i\) such that \(x_i\in
I_i\), and  choose a new reference  \(X' \defn \{x'_i\in I_i:  (-1)^i e(x'_i) =
\max_{x\in I_i} (-1)^i e(x)\}\).
\end{enumerate}

The Remez algorithm has several drawbacks:
\begin{itemize}
\item  It can  sometimes be  hard  to find  a suitable  initial reference:  the
location of  the zeros of Chebyshev  polynomials are sometimes  suggested as an
\emph{ad hoc} guess.
\item The  construction of the  interpolating rational function \(r\)  can fail
for some exceptional references.
\item Finding the  extrema of \(e(x)\) in an interval \(I_i\)  can be tricky if
the  function \(f\)  oscillates  rapidly  within the  interval.  As \(f\),  and
therefore  \(e\),  can be  any  continuous  function  an efficient  yet  robust
algorithm for locating the extrema is hard to find.
\item  The  computation often  has  to  be  carried out  in  multiple-precision
arithmetic: it is not unusual to need 20--30 or more decimal digits precision.
\item The rate of convergence is quite slow.
\end{itemize}
Of course, if bounds  on the spectrum of a family of  matrices is known \emph{a
priori} then  the cost of computing  the optimal rational  approximation is not
too  significant, as  it only  has to  be done  once for  many  matrix function
evaluations. This  is the case,  for example, in  computing roots of  the Dirac
operator for staggered fermions, whose  spectrum is bounded above by a constant
and below  by the mass. On  the other hand, in  the case of  Wilson fermions or
massless overlap Ginsparg-Wilson fermions a  useful lower bound on the spectrum
does  not  exist, and  one  either has  to  use  a higher-degree  approximation
covering a  conservative range or one  has to recompute  the approximation ``on
the fly,'' both of which may be prohibitively expensive options.

\section{Zolotarev's Theorem}

It  is fortunate  that  in the  most  interesting cases,  namely square  roots,
inverse square roots, and the sign function \(\sgn(x)\) the coefficients of the
optimal Chebyshev rational approximation are known analytically. This result is
due to Zolotarev, who was a student of Chebyshev.

Zolotarev  \cite{zolotarev:1877}  gave   an  explicit  expression  for  optimal
rational  approximations  to the  sign  function. To  be  precise,  he gave  an
expression of  the form \(r(x) =  xR(x^2)\) where \(R\) is  a rational function
and
\begin{equation}
  \max_{\epsilon \leq|x|\leq1}  |r(x) -  \sgn(x)| = \Delta.
  \label{eq:zolotarev-form}
\end{equation}
Note that  this degree \((n+1,n)\) approximation  is as near  to a ``diagonal''
rational  function  as possible,  since  \(r\)  must  be odd.  Furthermore,  no
approximation with  lower degree  numerator could be  better, for  otherwise it
would violate the uniqueness part of Chebyshev's theorem. 

\begin{figure}
  \epsfxsize=0.5\textwidth
  \centerline{\epspdffile{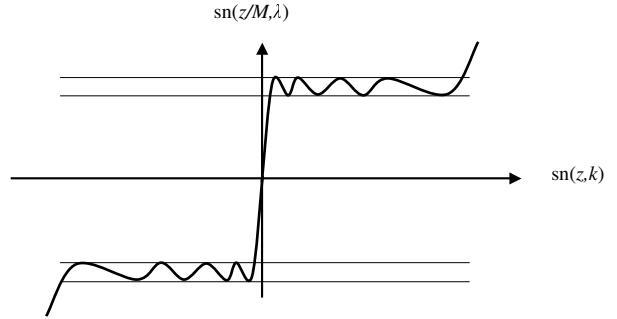}}
  \caption[Elliptic     2D]{Sketch    of     the    function     \(r(x)\)    of
    equations~(\ref{eq:zolotarev-form})  and~(\ref{eq:modular-transformation}).
    The four  values \(\pm1\pm\Delta\) are  indicated by the  horizontal lines,
    and \(r(x)\) crosses these lines at \(\pm\epsilon\) and \(\pm1\).}
  \label{fig:elliptic-2d}
\end{figure}

The approximation is illustrated in Figure~\ref{fig:elliptic-2d}.
It  is   clear  that  \(\tilde   r(x)  =  (1-\Delta^2)/r(x)\)  is   an  optimal
approximation to  \(\sgn(x)\) of degree  \((n,n+1)\), which is  singular rather
than  zero at  \(x=0\). If  there was  a better  approximation than  \(r\) with
higher degree numerator  then \(\tilde r\) would not  be unique, again contrary
to Chebyshev's theorem.

\subsection{Zolotarev Coefficients}

It is amusing that whereas  the properties of Chebyshev polynomials follow from
the fact that  \(\cos n\theta\) is a polynomial  in \(\cos\theta\), Zolotarev's
theorem follows  from an  analogous result for  elliptic functions.

The Jacobi elliptic function \(\sn\) is implicitly defined as the inverse of an
elliptic  integral,  namely \[z  \defn  \int_0^{\sn(z;k)} \!\!\!\!\!  \frac{dt}
{\sqrt{(1-t^2) (1-k^2t^2)}}.\] It is a \emph{doubly periodic} analytic function
in the  complex plane with periods  \(\omega=4\K(k)\) and \(\omega'=2i\K(k')\),
where \(\K\)  is the \emph{complete  elliptic integral} \[\K(k)  \defn \int_0^1
\frac{dt} {\sqrt{(1-t^2) (1-k^2t^2)}}\] and \(k^2 + k'^2 = 1\). It can be shown
\cite{achieser:1990,kennedy:2003b}  that  any   elliptic  function  with  these
periods  must be expressible  as a  rational function  of \(\sn(z;k)\)  and the
related  Jacobi  elliptic  functions  \(\cn(z;k)  =  \sqrt{1-\sn(z;k)^2}\)  and
\(\dn(z;k)  = \sqrt{1-k^2\sn(z;k)^2}\);  and  more specifically  that any  even
elliptic  function   with  the   same  periods  is   a  rational   function  of
\(\sn(z;k)^2\). In fact, any elliptic  function with equivalent periods, in the
sense that they generate the same period lattice, are rationally expressible in
this manner; as are elliptic functions with periods which divide \(\omega\) and
\(\omega'\), since the latter are also periodic with the original periods. Such
rational relationships are known as \emph{modular transformations}.

Consider  in particular  an  elliptic function  with  periods \(\tilde\omega  =
\omega  = 4\K(k)\)  and  \(\tilde\omega'  = \omega'/n  =  2i\K(k')/n\). Such  a
function is easily constructed from  \(\sn(z;k)\) by scaling the argument \(z\)
by some factor  \(1/M\) and choosing a new  parameter~\(\lambda\). The function
\(\sn(z/M;  \lambda)\)   has  periods  \(4L  =  4\K(\lambda)\)   and  \(2iL'  =
2i\K(\lambda')\)  where \(\lambda^2  + \lambda'^2  = 1\),  so the  periods with
respect to  \(z\) are \(4LM\)  and \(2iL'M\), which thus  requires \(LM=\K(k)\)
and \(L'M=\K(k')/n\). The  even elliptic function \(\sn(z/M;\lambda)/\sn(z;k)\)
must therefore be  expressible as a rational function  \(R\) of \(\sn(z;k)^2\),
\(\sn(z/M;\lambda) = \sn(z;k) R\left(\sn(z;k)^2\right)\).

Matching the  poles and  zeros, and fixing  the overall normalisation  from the
behaviour  near \(z=0\),  the  coefficients of  the  numerator and  denominator
polynomials for \(R\) may be expressed in terms of Jacobi elliptic functions,
\begin{equation}
  \frac{\sn(z/M;\lambda)}{\sn(z;k)} = \frac1M \prod_{m=1}^{\lfloor n/2\rfloor}
    \frac{1-\frac{\sn(z;k)^2}  {\sn(2iK'm/n;k)^2}}  {1-\frac{\sn(z;k)^2}
    {\sn\left(2iK'\left(m-\half\right)/n;k\right)^2}}.
  \label{eq:modular-transformation}
\end{equation}
The quantity \(M\) is determined by evaluating this identity at the half period
\(\K(k)\) where  \(\sn(\K(k);k) = \sn(\K(k)/M;\lambda) =  \sn(L;\lambda) = 1\).
Likewise, the parameter \(\lambda\) is  found by evaluating the identity at \(z
= \K(k) + i\K(k')/n\).

\begin{figure}
  \epsfxsize=0.5\textwidth
  \centerline{\epspdffile{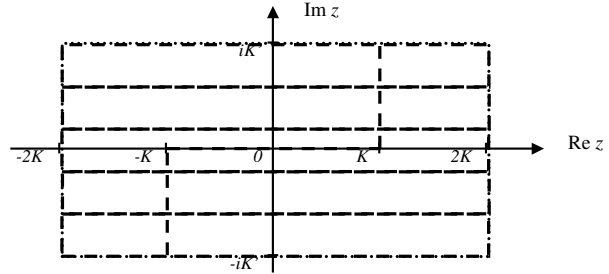}}
  \caption[fig:fundamental-regions]{The fundamental  region for \(\sn(z;k)\) is
    shown by the dotted box,  and that for \(\sn(z/M;\lambda)\) whose period is
    divided by \(n=5\)  along the imaginary axis is shown  by the dashed boxes.
    The line  from \(-K-iK'\)  through \(-K\) and  \(K\) to \(K+iK'\) shows the
    contour in the complex \(z\) plane along which both functions are real, and
    \(\sn(z;k)\)   takes   values  from   \(-1\)   through  \(-\epsilon\)   and
    \(\epsilon\) to \(1\)  while  \(\sn(z/M;\lambda)\) oscillates about it with
    amplitude \(\Delta\).}
  \label{fig:fundamental-regions}
\end{figure}

\begin{figure}
  \epsfxsize=0.5\textwidth
  \centerline{\epspdffile{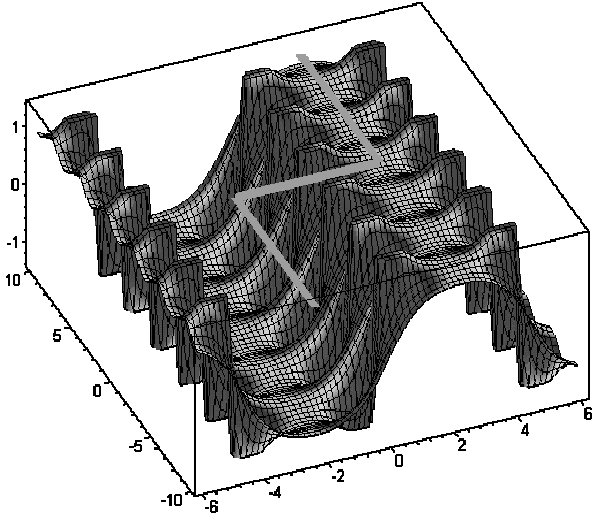}}
  \caption[Elliptic    3D]{Surface   plot    showing   the    real    part   of
    \(\sn(z/M;\lambda)\)  over  the  fundamental  region of  \(\sn(z;k)\).  The
    imaginary part vanishes along the contour indicated.}
  \label{fig:elliptic-3d}
\end{figure}

The fundamental  region for  \(\sn(z;k)\) and \(\sn(z/M;\lambda)\)  for \(n=5\)
are shown in Figure~\ref{fig:fundamental-regions}, which also shows the contour
in the  complex \(z\)  plane along which  the modular transformation  gives the
desired real rational relationship between these two elliptic functions.
Some  further insight is  given by  Figure~\ref{fig:elliptic-3d}, in  which the
real part of \(\sn(z/M;\lambda)\) is shown  over the same region of the complex
plane.   The  imaginary   part  vanishes   along  the   contour   described  in
Figure~\ref{fig:fundamental-regions},  which is  also indicated  on  this plot.
From  this plot  it is  clear how  the two  periods of  the  elliptic functions
correspond to the  ``step'' of the sign function and  the oscillations of \(r\)
about it.

\subsection{Approximations to $\sqrt x$ and $1/\sqrt x$}

We may rewrite equation~(\ref{eq:zolotarev-form}) in the form
\begin{eqnarray*}
  \Delta &=& \max_{\epsilon \leq|x|\leq1}\left|xR(x^2) - \frac x{|x|}\right| \\
  &=& \max_{\epsilon \leq|x|\leq1} \left|R(x^2) - \frac1{|x|}\right|\cdot|x|   
\end{eqnarray*}
and,   upon   setting   \(\xi=x^2\),   we   find  the   relation   \(\Delta   =
\max_{\sqrt\epsilon \leq\xi\leq1} \left|R(\xi) - \frac1{\sqrt\xi} \right| \cdot
\sqrt\xi\). This shows  that \(R(x)\) is an optimal  Chebyshev approximation to
the  function  \(1/\sqrt  x\)  with   weight  \(\sqrt  x\)  over  the  interval
\([\sqrt\epsilon,1]\), or  in other  words the optimal  Chebyshev approximation
with unform  {\sl relative} error \(\Delta  = \max_{\sqrt\epsilon \leq\xi\leq1}
\frac{\left|R(\xi) - 1/\sqrt\xi\right|} {1/\sqrt\xi}\).  We can rewrite this as
\(\Delta   =  \max_{\sqrt\epsilon   \leq\xi\leq1}   \frac{\left|\xi  R(\xi)   -
\sqrt\xi\right|}{\sqrt\xi}\),   so   \(xR(x)\)    is   an   optimal   Chebyshev
approximation to  \(\sqrt x\) over \([\sqrt\epsilon,1]\)  with uniform relative
error.

\section{Examples}

\subsection{A Simple Numerical Example}

It might be useful to look at an actual example of the optimal minimax rational
approximations we have been discussing. The optimal degree \((3,3)\)
approximation to \(1/\sqrt x\) over the interval \([0.003,1]\) is
\begin{eqnarray*}
  \frac1{\sqrt x} &\approx& 0.3904603901
    \frac{(x+2.3475661045)}{(x+0.4105999719)} \times \\
    &\times& \frac{(x+0.1058344600) (x+0.0073063814)} 
      {(x+0.0286165446) (x+0.0012779193)},
\end{eqnarray*}
and, as promised, it has a numerically stable partial fraction expansion
\begin{eqnarray*}
  \frac1{\sqrt x} &\approx& 0.3904603901 + 
    \frac{0.0511093775}{x+0.0012779193} + \\
    &+& \frac{0.1408286237}{x+0.0286165446} +
    \frac{0.5964845033}{x+0.4105999719}.
\end{eqnarray*}
Notice that all the terms are positive, and the roots of the denominators are
all negative.

\subsection{Comparison of Polynomial and Rational Approximations}

We now briefly  compare the quality of polynomial  and rational approximations.
It  is  common folklore  that  even  though  they require  division  operations
rational approximations are  superior to polynomial ones, and  this seems to be
true for  matrix function  approximations too,  if use is  made of  the partial
fraction expansion and  multi-shift solver tricks. In a  sense, as we mentioned
before, this is because Krylov space solvers are more effective than polynomial
approximations to  \(1/x\) of  fixed degree  as a means  of solving  systems of
linear equations.

\begin{figure}
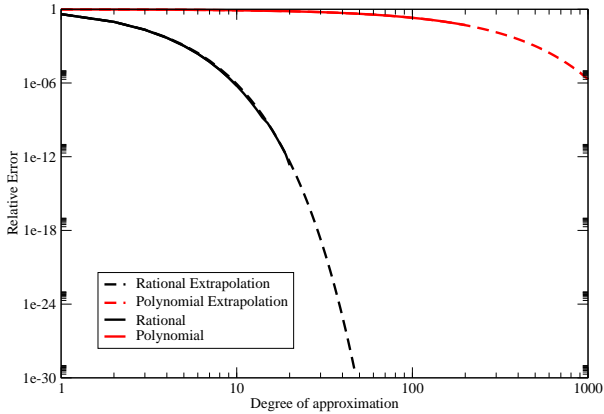

  \epsfxsize=0.5\textwidth
  \centerline{\epspdffile{degree}}
  \caption[Rational  Error]{Comparison of minimax  errors for  optimal rational
    and polynomial  approximating functions  to \(x^{-1/2}\) over  the interval
    [0.00003,1]   (corresponding  to   a  staggered   fermion   mass  parameter
    \(m=0.025\)) as a function of approximation degree.}
  \label{fig:rat-error}
\end{figure}

A  numerical  comparison  of  the  minimax  errors  for  optimal  rational  and
polynomial  approximating   functions  to  \(1/\sqrt  x\)   over  the  interval
\([0.00003,1]\)  (corresponding  to  a  staggered  fermion  mass  parameter  of
\(m=0.025\))   as   a   function   of   approximation  degree   is   shown   in
Figure~\ref{fig:rat-error}. 
The key points to note are:
\begin{itemize}
\item The error  of the rational approximations falls  exponentially with their
degree.  Machine-precision  errors of  one  \emph{ulp}  (about \(10^{-7}\)  for
32-bit IEEE  floating-point arithmetic and \(10^{-15}\)  for 64-bit arithmetic)
are easily achieved for relatively low degree rational functions.
\item In order to reach small  errors, of the scale of the floating-point error
scale  of  one \emph{ulp},  using  polynomials  requires  extemely high  degree
polynomials. Considerable care needs to be taken with regard to rounding errors
in these cases.
\item  The amount  of work  needed to  apply the  rational approximations  of a
sparse matrix to  a vector, using a partial  fraction expansion and multi-shift
solver, depends  essentially on the degree  of the rational  function times the
dimension of the Krylov space required.
\end{itemize}

\begin{figure}
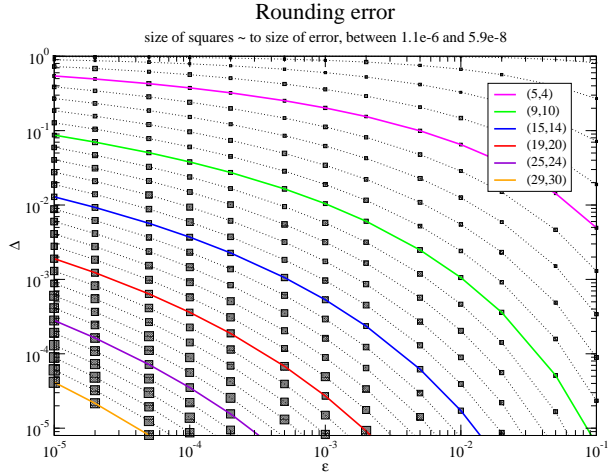

  \epsfxsize=0.5\textwidth
  \centerline{\epspdffile{eps-delta}}
  \caption[Zolotarev  Errors]{The   minimax  error  \(\Delta\)   for  Zolotarev
    approximations of degree \(n\) over the interval \([\epsilon,1]\). The size
    of the squares indicates the size  of the rounding errors in evaluating the
    Zolotarev rational functions in single-precision arithmetic.}
  \label{fig:zolotarev-errors}
\end{figure}

The  dependence of  the minimax  error  \(\Delta\) for  the Zolotarev  rational
approximations on  the approximation  interval \([\epsilon,1]\) and  the degree
\(n\) is shown in Figure~\ref{fig:zolotarev-errors}. This plot also attempts to
indicate  the   numerical  errors  in  evaluating  the   rational  function  in
single-precision  arithmetic by  the  size  of the  squares.  Again, the  rapid
convergence  and  stability  of  the  approximations is  manifest.  The  errors
\(\Delta\)  appear to  be consistent  with an  asymptotic formula  of  the form
\(\Delta \propto e^{n/\ln\epsilon}\)~\cite{petrushev:1987}.

At the end of section~\ref{sec:remez} we noted that it was expensive to use the
Remez algorithm  to compute optimal  rational approximations ``on the  fly'' in
the case where we do not have a  useful lower bound on the spectrum of a family
of  Dirac operators.  For the  interesting cases  where Zolotarev's  formula is
applicable  we  can compute  the  coefficients  in  the rational  approximation
efficiently ``on the fly'' if we  measure the smallest eigenvalue of each Dirac
operator.  The  method of  computing  the  Zolotarev  coefficients rapidly  and
accurately  making  use  of  Gauss'  arithmetico-geometric  mean  is  explained
in~\cite{kennedy:2003b}.

\section{Conclusion}

We reviewed the subject of minimax  approximation theory, and explain why it is
particularly  effective   for  the   approximation  of  matrix   functions.  We
demonstrate why rational approximations  are often superior to polynomial ones,
and how the  combination of partial fraction expansions  and multi-shift Krylov
space solvers allows  them to be applied efficiently  to large sparse matrices.
Finally, we have  outlined how the optimal approximation  for the sign function
may be found in closed form using elliptic functions.

\section*{Acknowledgments}

I   would  like   to  thank   Urs  Wenger   and  Mike   Clark  for   help  with
Figures~\ref{fig:zolotarev-errors}  and~\ref{fig:rat-error}  respectively,  and
Ken Bowler for useful comments. I would also like to thank CSSM for inviting me
to the very pleasant and productive workshop in Cairns.

\end{document}